\pdfoutput=1

\documentclass[11pt,twoside,a4paper,cmspaper,final,collab]{cms-tdr}
\def\svnVersion{52412:55246MP}\def\cmsCernNo{2011-056}\def\cmsCernDate{\today}\def\cmsMessage{Submitted to the Journal of High Energy Physics}

\begin{document}\cmsNoteHeader{HIN-11-001}

\hyphenation{had-ron-i-za-tion}
\hyphenation{cal-or-i-me-ter}
\hyphenation{de-vices}
\RCS$Revision: 55246 $
\RCS$HeadURL: svn+ssh://alverson@svn.cern.ch/reps/tdr2/papers/HIN-11-001/trunk/HIN-11-001.tex $
\RCS$Id: HIN-11-001.tex 55246 2011-05-12 11:12:32Z alverson $
\newcommand {\roots}    {\ensuremath{\sqrt{s}}}
\newcommand {\rootsNN}  {\ensuremath{\sqrt{s_{_{NN}}}}}
\newcommand {\dndy}     {\ensuremath{dN/dy}}
\newcommand {\dnchdy}   {\ensuremath{dN_{\mathrm{ch}}/dy}}
\newcommand {\dndeta}   {\ensuremath{dN/d\eta}}
\newcommand {\dnchdeta} {\ensuremath{dN_{\mathrm{ch}}/d\eta}}
\newcommand {\dndpt}    {\ensuremath{dN/d\pt}}
\newcommand {\dnchdpt}  {\ensuremath{dN_{\mathrm{ch}}/d\pt}}
\newcommand {\deta}     {\ensuremath{\Delta\eta}}
\newcommand {\dphi}     {\ensuremath{\Delta\phi}}
\newcommand {\pttrg}       {\ensuremath{p_\mathrm{T}^{\mathrm{trig}}}}
\newcommand {\ptass}       {\ensuremath{p_\mathrm{T}^{\mathrm{assoc}}}}
\newcommand {\AJ}       {\ensuremath{A_J}}
\newcommand {\npart}    {\ensuremath{N_{part}}}
\newcommand {\ncoll}    {\ensuremath{N_{coll}}}

\newcommand {\pp}    {\mbox{pp}}
\newcommand {\ppbar} {\mbox{p\={p}}}
\newcommand {\pbarp} {\mbox{p\={p}}}
\newcommand {\PbPb}  {\mbox{PbPb }}

\newcommand{\m}{\ensuremath{\,\text{m}}\xspace}
\newcommand {\naive}    {na\"{\i}ve}
\providecommand{\GEANT} {{Geant}\xspace}
\providecommand{\PHOJET} {\textsc{phojet}\xspace}
\providecommand{\PYNEW} {\textsc{pythia8}\xspace}

\def\d{\mathrm{d}}

\providecommand{\PKzS}{\ensuremath{\mathrm{K^0_S}}}
\providecommand{\Pp}{\ensuremath{\mathrm{p}}}
\providecommand{\Pap}{\ensuremath{\mathrm{\overline{p}}}}
\providecommand{\PgL}{\ensuremath{\Lambda}}
\providecommand{\PagL}{\ensuremath{\overline{\Lambda}}}
\providecommand{\PgS}{\ensuremath{\Sigma}}
\providecommand{\PgSm}{\ensuremath{\Sigma^-}}
\providecommand{\PgSp}{\ensuremath{\Sigma^+}}
\providecommand{\PagSm}{\ensuremath{\overline{\Sigma}^-}}
\providecommand{\PagSp}{\ensuremath{\overline{\Sigma}^+}}
\renewcommand{\GEANTfour} {{Geant4}\xspace}

\cmsNoteHeader{HIN-11-001} 
\title{Long-range and short-range dihadron angular correlations in central PbPb collisions
at \rootsNN\ = 2.76\TeV}

\date{\today}

\abstract{
First measurements of dihadron correlations for charged particles are presented
for central PbPb collisions at a nucleon-nucleon center-of-mass
energy of 2.76\TeV over a broad range in relative pseudorapidity ($\Delta\eta$)
and the full range of relative azimuthal angle ($\Delta\phi$).
The data were collected with the CMS detector, at the LHC. A broadening
of the away-side ($\Delta\phi \approx \pi$) azimuthal correlation is
observed at all $\Delta\eta$, as compared to the measurements in
pp collisions. Furthermore, long-range dihadron
correlations in $\Delta\eta$ are observed for
particles with similar $\phi$ values. This phenomenon, also
known as the ``ridge'', persists up to at least $|\Delta\eta| = 4$.
For particles with transverse momenta (\pt) of 2--4\GeVc, the
ridge is found to be most prominent when these particles are
correlated with particles of \pt\ = 2--6\GeVc, and to be much
reduced when paired with particles of \pt\ = 10--12\GeVc.
}

\hypersetup{%
pdfauthor={CMS Collaboration},%
pdftitle={Long-range and short-range dihadron angular correlations
in central PbPb collisions at a nucleon-nucleon center of mass energy of 2.76 TeV},%
pdfsubject={CMS},%
pdfkeywords={CMS, physics}}

\maketitle 

\section{Introduction}

Measurements of dihadron azimuthal correlations~\cite{Adler:2002tq,star:2009qa,Adams:2005ph,Adare:2006nr,phenix:2008cqb,
Alver:2009id,Alver:2008gk} have provided a powerful
tool to study the properties of the strongly interacting
medium created in ultrarelativistic nuclear
collisions~\cite{Adams:2005dq,Adcox:2004mh,Back:2004je,Arsene:2004fa}.
An early indication of strong jet-medium interactions at RHIC was
the absence of high-transverse-momentum (high-\pt) back-to-back
particle pairs in dihadron correlation measurements~\cite{Adler:2002tq}
and the corresponding enhancement of low-\pt hadrons recoiling from a high-\pt
leading, or ``trigger'', particle~\cite{Adams:2005ph}. The recent observations of
the suppression of high-\pt charged hadrons~\cite{Aamodt:2010jd} and of asymmetric
energies of reconstructed jets~\cite{Aad:2010bu,HIN-10-004} in PbPb collisions
at the Large Hadron Collider (LHC) provide further evidence of jet quenching,
suggesting a large energy loss for partons traversing the produced medium.

At RHIC, extending dihadron azimuthal correlation measurements to larger
relative pseudorapidities resulted in the discovery of a ridge-shaped correlation
in central AuAu collisions between particles with small relative azimuthal
angles ($|\Delta\phi| \approx 0$), out to very large relative pseudorapidity
($|\Delta\eta|$)~\cite{star:2009qa,Alver:2009id}. Although the ``ridge''
has been qualitatively described by several different models~\cite{Armesto:2004pt,
Majumder:2006wi,Chiu:2005ad,Wong:2008yh,Voloshin:2003ud,Romatschke:2006bb,
Shuryak:2007fu,Dumitru:2008wn,Gavin:2008ev,Dusling:2009ar,Hama:2009vu,Alver:2010gr},
its origin is still not well understood. Some
models attribute the ridge to jet-medium interactions,
while others attribute it to the medium itself. The ridge has
been observed for particles with transverse momenta from several hundred \!\MeVc to a few \!\GeVc.
However, the character of the ridge for even higher-\pt
particles, as well as its dependence on collision energy, is still
poorly understood from the RHIC results~\cite{star:2009qa}.
Recently, a striking ridge structure has also been observed in very high
multiplicity proton-proton (pp) collisions at a center-of-mass energy of 7\TeV
at the LHC by the Compact Muon Solenoid (CMS)
Collaboration~\cite{Khachatryan:2010gv}, posing new challenges to the
understanding of these long-range correlations.

This paper presents the first measurement of dihadron correlations
for charged particles produced in the most central (0--5\%
centrality) PbPb collisions at a nucleon-nucleon center-of-mass energy
(\rootsNN) of 2.76\TeV over a large phase space. The results are
presented in terms of the associated hadron yields as a function
of pseudorapidity and azimuthal angle relative to trigger
particles in different transverse momentum intervals.
Traditionally, trigger particles have been utilized to represent
the direction of the leading hadron in a jet, and were
required to have a higher momentum than all the other associated particles in the jet~\cite{star:2009qa,Alver:2009id}.
However, as shown in Ref.~\cite{Khachatryan:2010gv},
important information can also be obtained by studying
the correlation of hadron pairs from the same \pt interval, which is
particularly useful when addressing the properties of the medium itself.
The current analysis employs both approaches. This measurement provides
a unique examination of the ridge in the most central PbPb
collisions at the highest energies reached so far in the laboratory
over a wide range in transverse
momentum (2--12\GeVc) and up to large relative
pseudorapidity ($|\Delta\eta| \approx 4$), imposing further
quantitative constraints on the possible origin of the ridge.

Details of event readout and analysis for extracting
the correlation functions are described in Section~\ref{sec:data},
the physics results found using the correlations are described
in Section~\ref{sec:results}, and a summary is given in
Section~\ref{sec:conclusions}.

\section{Data and Analysis}
\label{sec:data}
The analysis reported in this paper is based on PbPb collisions
at \rootsNN\ = 2.76\TeV collected during the LHC heavy-ion run
in November and December 2010 with the CMS detector.
The central feature of the CMS apparatus is a superconducting solenoid
of 6~m internal diameter. Within the field volume are the inner
tracker, the crystal electromagnetic calorimeter,
and the brass/scintillator hadron calorimeter.
Muons are measured in gas-ionization detectors embedded in the
steel return yoke. In addition to the barrel and endcap detectors,
CMS has extensive forward calorimetry.
The nearly 4$\pi$ solid-angle
acceptance of the CMS detector is ideally suited for studies of both short-
and long-range particle correlations. A detailed description
of the CMS detector can be found in Ref.~\cite{JINST}. CMS uses a
right-handed coordinate system, with the origin at the
nominal interaction point, the $x$ axis pointing to the center of the
LHC, the $y$ axis pointing up (perpendicular to the LHC plane), and
the $z$ axis along the counterclockwise beam direction.
The detector subsystem primarily used for the present analysis is
the inner tracker that reconstructs the trajectories of charged particles
with $\pt > 100$\MeVc,
covering the pseudorapidity region $|\eta| < 2.5$, where
$\eta = -\ln [ \tan(\theta/2)]$ and $\theta$ is
the polar angle relative to the beam direction.
The inner tracker consists of 1440 silicon pixel and 15\,148 silicon strip
detector modules immersed in the 3.8~T axial magnetic field of the
superconducting solenoid.

The event readout of the CMS detector for PbPb collisions is triggered by
coincident signals in forward detectors located on both sides of the nominal
collision point. In particular, minimum bias PbPb data are recorded based on
coincident signals in the beam scintillator
counters (BSC, $3.23 < |\eta| < 4.65$) or in the steel/quartz-fiber
Cherenkov forward hadron calorimeters (HF, $2.9 < |\eta| < 5.2$)
from both ends of the detector. In order to suppress events
due to noise, cosmic rays, double-firing triggers, and beam backgrounds,
the minimum bias trigger used in this analysis is required
to be in coincidence with bunches colliding in the interaction region.
The trigger has an acceptance of $(97 \pm 3)$\%
for hadronic inelastic PbPb collisions~\cite{HIN-10-004}.

Events are selected offline by requiring in addition at least three hits
in the HF calorimeters at both ends of CMS, with at least
3\GeV of energy in each cluster, and the presence of a reconstructed
primary vertex containing at least two tracks. These criteria further reduce
background from single-beam interactions (e.g., beam gas and beam halo),
cosmic muons, and large-impact-parameter,
ultra-peripheral collisions that lead to the electromagnetic
breakup of one or both of the Pb nuclei~\cite{Djuvsland:2010qs}.
The reconstructed primary vertex is required to be located
within 15~cm of the nominal collision point along the beam axis
and within a radius of 0.02~cm relative to the average vertex position
in the transverse plane.

This analysis is based on a data sample of PbPb collisions corresponding
to an integrated luminosity of approximately 3.12~$\mu$b$^{-1}$~\cite{EWK-10-004,EWK-11-001},
which contains 24.1~million minimum bias
collisions after all event selections are applied.

The energy released in the collisions is related to the centrality
of heavy-ion interactions, i.e., the geometrical overlap
of the incoming nuclei. In CMS, centrality is classified according to percentiles of the
distribution of the energy deposited in the HF calorimeters. The centrality class
used in this analysis corresponds to the 0--5\% most central PbPb
collisions, a total of 1.2~million events. More details on the centrality
determination can be found in Refs.~\cite{HIN-10-004,D'Enterria:2007xr}.

A reconstructed track is considered as a primary-track candidate if
the significance of the separation along the beam axis between
the track and the primary vertex, $d_{z}/\sigma(d_{z})$, and the significance
of the impact parameter relative to the primary vertex transverse to the beam,
$d_{xy}/\sigma(d_{xy})$, are each less than 3. In order to remove tracks
with potentially poorly reconstructed momentum values, the relative uncertainty
of the momentum measurement, $\sigma(\pt)/\pt$, is required to be less than 5.0\%.
Requiring at least 12 hits on each track helps to reject
misidentified tracks. Systematic uncertainties related to the
track selections have been evaluated as discussed below.

Trigger particles are defined as all charged particles originating
from the primary vertex, with $|\eta| < 2.4$ and in a specified \pttrg\ range.
The number of trigger particles in the event is denoted by $N_{\rm trig}$,
which can be more than one per event.
Hadron pairs are formed by associating with every trigger particle
the remaining charged particles with $|\eta| < 2.4$ and in a specified \ptass\
range. The per-trigger-particle associated yield distribution is then defined by:

\vspace{-0.2cm}
\begin{equation}
\label{2pcorr_incl}
\frac{1}{N_{\rm trig}}\frac{d^{2}N^{\rm pair}}{d\Delta\eta d\Delta\phi}
= B(0,0)\times\frac{S(\Delta\eta,\Delta\phi)}{B(\Delta\eta,\Delta\phi)},
\end{equation}
\vspace{-0.2cm}

\noindent where $\Delta\eta$ and $\Delta\phi$ are the differences in $\eta$
and $\phi$ of the pair, respectively. The signal distribution, $S(\Delta\eta,\Delta\phi)$, is
the measured per-trigger-particle distribution of same-event pairs, i.e.,

\vspace{-0.2cm}
\begin{equation}
\label{eq:signal}
S(\Delta\eta,\Delta\phi) = \frac{1}{N_{\rm trig}}\frac{d^{2}N^{\rm same}}{d\Delta\eta d\Delta\phi}.
\end{equation}
\vspace{-0.2cm}

\noindent
The mixed-event background distribution,

\vspace{-0.2cm}
\begin{equation}
\label{eq:background}
B(\Delta\eta,\Delta\phi) = \frac{1}{N_{\rm trig}}\frac{d^{2}N^{\rm mix}}{d\Delta\eta d\Delta\phi},
\end{equation}
\vspace{-0.2cm}

\noindent is constructed by pairing the trigger particles in each event with the
associated particles from 10 different random events, excluding
the original event. The symbol $N^{\rm mix}$ denotes the number of
pairs taken from the mixed event. The background distribution is used to account
for random combinatorial background and pair-acceptance effects.
The normalization factor $B(0,0)$ is
the value of $B(\Delta\eta,\Delta\phi)$ at $\Delta\eta=0$ and $\Delta\phi=0$
(with a bin width of 0.3 in $\Delta\eta$ and $\pi/16$ in $\Delta\phi$), representing
the mixed-event associated yield for both particles of the pair
going in approximately the same direction,
thus having full pair acceptance.
Therefore, the ratio $B(0,0)/B(\Delta\eta,\Delta\phi)$
is the pair-acceptance correction factor used to derive the corrected
per-trigger-particle associated yield distribution. Equation~(\ref{2pcorr_incl}) is calculated
in 2~cm wide bins of the vertex position ($z_{\rm vtx}$)
along the beam direction and averaged over the range
$|z_{\rm vtx}| < 15$~cm. To maximize the statistical precision,
the absolute values of $\Delta\eta$ and $\Delta\phi$ are used to fill one quadrant of
the ($\Delta\eta,\Delta\phi$) histograms, with the other three quadrants
filled (only for illustration purposes) by
reflection. Therefore, the resulting distributions are symmetric about
$(\Delta\eta,\Delta\phi) = (0,0)$ by construction.

Each reconstructed track is weighted by the inverse of the
efficiency factor, $\varepsilon_{\rm trk}(\eta,\pt)$,
as a function of the track's pseudorapidity and transverse momentum.
The efficiency weighting factor accounts for the detector
acceptance $A(\eta,\pt)$, the reconstruction efficiency $E(\eta,\pt)$,
and the fraction of misidentified tracks, $F(\eta,\pt)$,

\vspace{-0.2cm}
\begin{equation}
\varepsilon_{\rm trk}(\eta,\pt) = \frac{A E}{1-F}.
\end{equation}
\vspace{-0.2cm}

\noindent Studies with simulated Monte Carlo (MC) events show that the combined geometrical acceptance
and reconstruction efficiency for the primary-track reconstruction
reaches about 60\% for the 0--5\% most central PbPb collisions at $\pt > 2\GeVc$
over the full CMS tracker acceptance ($|\eta| < 2.4$) and 65\% for $|\eta| < 1.0$.
The fraction of misidentified tracks is about 1--2\%
for $|\eta| < 1.0$, but increases to 10\% at $|\eta| \approx 2.4$.
The weighting changes the overall scale but not the shape of the associated
yield distribution, which depends on the ratio of the signal to background
distributions.

A closure test of the track-weighting procedure is performed
on {\sc HYDJET}~\cite{Lokhtin:2005px} (version 1.6) MC events.
The efficiency-weighted associated yield distribution from reconstructed tracks is
found to agree with the generator-level correlation function to within 3.3\%.
In addition, systematic checks of the tracking efficiency, in which simulated
MC tracks are embedded into data events, give results consistent
with pure {\sc HYDJET} simulations to within 5.0\%. The tracking
efficiency also depends on the vertex $z$ position of the event.
However, this dependence is negligible in this analysis, and
its effects are taken into account in the systematic uncertainty
by comparing the efficiency-corrected correlation functions
for two different $z_{\rm vtx}$ ranges,
$|z_{\rm vtx}| < 15\cm$ and $|z_{\rm vtx}| < 5\cm$,
which are found to differ by less than 2.2\%. Additional uncertainties due to track
quality cuts are examined by loosening or tightening
the track selections described previously, and the final results are found to be insensitive to
the selections to within 2.0\%. An independent analysis,
using a somewhat different but well-established methodology~\cite{Aggarwal:2010rf,Chetluru:2011}
in constructing the mixed-event background is performed as a
cross-check, where 10 trigger particles from different events are selected first
and combined to form a single event, and then correlated with particles
from another event. It yields results within 2.9--3.6\%
of the default values (3.6\% for $|\Delta\eta| < 1$ and 2.9\%
for $2 < |\Delta\eta| < 4$), with a slight dependence on $\Delta\eta$
and $\Delta\phi$. The other four sources of systematic uncertainty
are largely independent of $\Delta\eta$ and $\Delta\phi$.

Table~\ref{tab:syst-table-new} summarizes the different systematic
sources, whose corresponding uncertainties are added in quadrature
becoming the quoted systematic uncertainties of the
per-trigger-particle associated yield.

\begin{table}[ht]
\caption{\label{tab:syst-table-new} Summary of systematic uncertainties.}

\begin{center}
\begin{tabular}{lc}
\hline
\hline
 Source                                          & Systematic uncertainty of \\
 & the per-trigger-particle associated yield (\%)  \\
\hline
 Tracking weighting closure test                                     & 3.3 \\
 Tracking efficiency                                                  & 5.0 \\
 Vertex dependence                                                    & 2.2 \\
 Track selection dependence                                           & 2.0 \\
 Construction of the mixed-event background                                & 2.9--3.6 \\
\hline
 Total                                                                & 7.3--7.6 \\
\hline
\hline
\end{tabular}
\end{center}
\end{table}

\section{Results}
\label{sec:results}

\begin{figure}[thb]
  \begin{center}
    \subfigure{\includegraphics[width=0.49\linewidth]{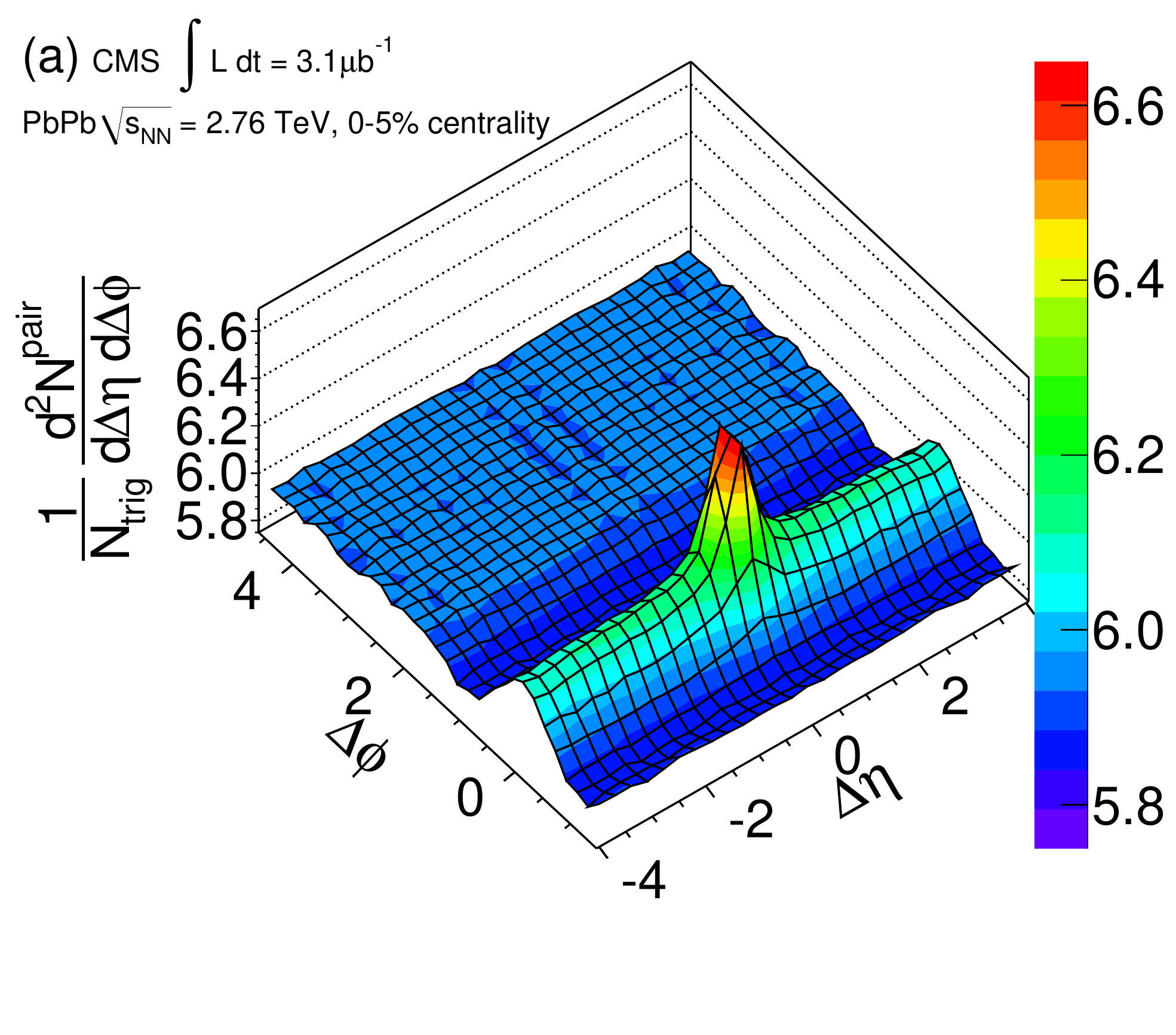} }
    \subfigure{\includegraphics[width=0.49\linewidth]{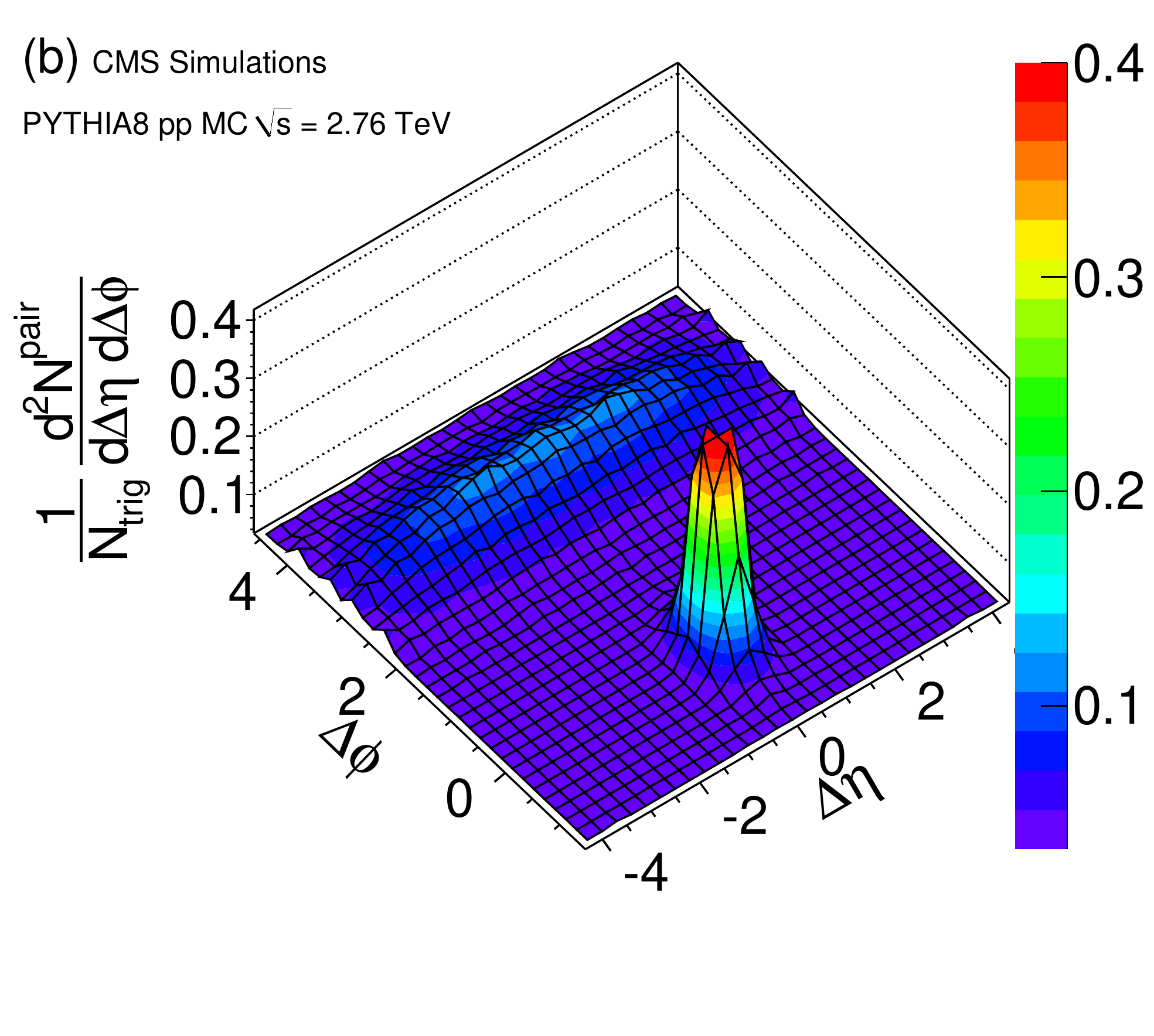} }
    \caption{
         Two-dimensional (2-D) per-trigger-particle associated yield of charged hadrons
         as a function of $|\Delta\eta|$ and $|\Delta\phi|$ for
         $4 < \pttrg < 6$\GeVc and $2 < \ptass < 4$\GeVc from
         (a) 0--5\% most central PbPb collisions at \rootsNN\ = 2.76\TeV,
         and (b) \PYNEW pp MC simulation at \roots\ = 2.76\TeV.}
    \label{fig:corr2D_final}
  \end{center}
\end{figure}

The measured per-trigger-particle associated yield distribution of
charged hadrons as a function of $|\Delta\eta|$ and $|\Delta\phi|$
in the 0--5\% most central PbPb collisions at \rootsNN\ = 2.76\TeV
is shown in Fig.~\ref{fig:corr2D_final}a for trigger particles with
$4 < \pttrg < 6\GeVc$ and associated particles with
$2 < \ptass < 4\GeVc$. To understand the effects of the hot, dense medium produced in
the collisions, this distribution can be compared to that from a
\PYNEW MC simulation~\cite{Sjostrand:2007gs} (version 8.135) of
pp collisions at \roots\ = 2.76\TeV, shown in Fig.~\ref{fig:corr2D_final}b.
This transverse momentum range, one of several studied later in this
paper, is chosen for this figure since it illustrates the differences
between correlations from PbPb data and \PYNEW pp MC events most clearly.
The main features of the simulated pp MC distribution are
a narrow jet-fragmentation peak at $(\Delta\eta, \Delta\phi) \approx (0, 0)$
and a back-to-back jet structure at $|\Delta\phi| = \pi$, but extended in $\Delta\eta$.
In the 0--5\% most central PbPb collisions, particle correlations
are significantly modified, as shown
in Fig.~\ref{fig:corr2D_final}a. The away-side pairs ($\Delta\phi \approx \pi$)
exhibit a correlation with similar amplitude compared to \PYNEW, although the structure in PbPb data
is much broader in both $\Delta\phi$ and $\Delta\eta$ so that it appears almost flat.
On the near side ($\Delta\phi \approx 0$), besides the common jet-like
particle production in both pp and PbPb at $(\Delta\eta, \Delta\phi) \approx (0, 0)$
due to jet fragmentation, a clear and significant ridge-like
structure is observed in PbPb at $\Delta\phi \approx 0$, which extends
all the way to the limit of the measurement of $|\Delta\eta| = 4$.

In relativistic
heavy-ion collisions, the anisotropic hydrodynamic expansion of the produced medium
is one possible source of long-range azimuthal correlations, driven
by the event-by-event initial anisotropy of the collision zone~\cite{Alver:2006wh,Alver:2008gk}.
For non-central collisions, these correlations are dominated by the second-order
Fourier component of the $|\Delta\phi|$ distribution, usually called elliptic flow or $v_2$.
Measurements of dihadron correlations at RHIC have frequently attempted to subtract
or factorize the elliptic flow contribution based on
direct $v_{2}$ measurements, in order to reveal
other features of particle correlations that may provide insight into
the interactions between the jets and the medium. However, recent
theoretical developments indicate that the interplay between
initial-state fluctuations and the subsequent hydrodynamic expansion
gives rise to additional Fourier components in the azimuthal
particle correlations~\cite{Hama:2009vu,Alver:2010gr,
Alver:2010dn,Schenke:2010rr,Petersen:2010cw,Xu:2010du,Teaney:2010vd}.
These components need to be treated on equal footing with the elliptic flow component.
In particular for the 0--5\% most central PbPb collisions, the elliptic flow
contribution to the azimuthal correlations
is not expected to be dominant~\cite{Aamodt:2010pa}.
Therefore, the original unsubtracted correlation functions are
presented in this paper, containing the full information necessary
for the comparison with theoretical calculations.

\subsection{Associated Yield Distributions versus $\Delta\phi$}

To quantitatively examine the features of short-range and long-range
azimuthal correlations, one dimensional (1-D) $\Delta\phi$
correlation functions are calculated by averaging the 2-D distributions
over a limited region in $\Delta\eta$ from $\Delta\eta_{\rm min}$ to $\Delta\eta_{\rm max}$:

\vspace{-0.2cm}
\begin{equation}
\label{2pcorr_incl_1D}
\frac{1}{N_{\rm trig}}\frac{dN^{\rm pair}}{d\Delta\phi}
= \frac{1}{\Delta\eta_{\rm max}-\Delta\eta_{\rm min}}
\int_{\Delta\eta_{\rm min}}^{\Delta\eta_{\rm max}}
\frac{1}{N_{\rm trig}}\frac{d^{2}N^{\rm pair}}{d\Delta\eta d\Delta\phi}d\Delta\eta.
\end{equation}
\vspace{-0.2cm}

\begin{figure}[thb]
  \begin{center}
    \includegraphics[width=1.0\textwidth]{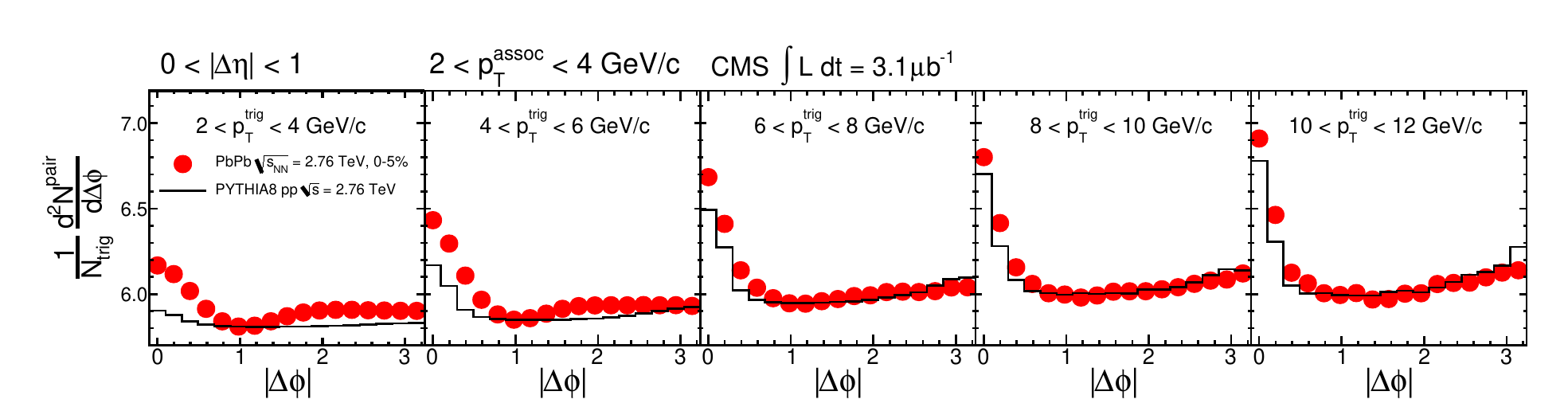}
    \caption{Short-range ($|\Delta\eta| < 1$)
        per-trigger-particle associated yields of charged hadrons as a function
        of $|\Delta\phi|$ from the 0--5\% most central PbPb collisions at \rootsNN\ = 2.76\TeV,
        requiring $2 < \ptass < 4$\GeVc, for five different intervals of \pttrg.
        The \PYNEW pp MC results (solid histograms) are also shown, shifted up
        by a constant value to match the PbPb data at $\Delta\phi=1$
        for ease of comparison. The error bars are statistical only
        and are too small to be visible in most of the panels. The systematic uncertainty
        of 7.6\% for all data points is not shown in the plots.}
    \label{fig:corr1D_eta1}
  \end{center}
\end{figure}

\begin{figure}[thb]
  \begin{center}
    \includegraphics[width=1.0\textwidth]{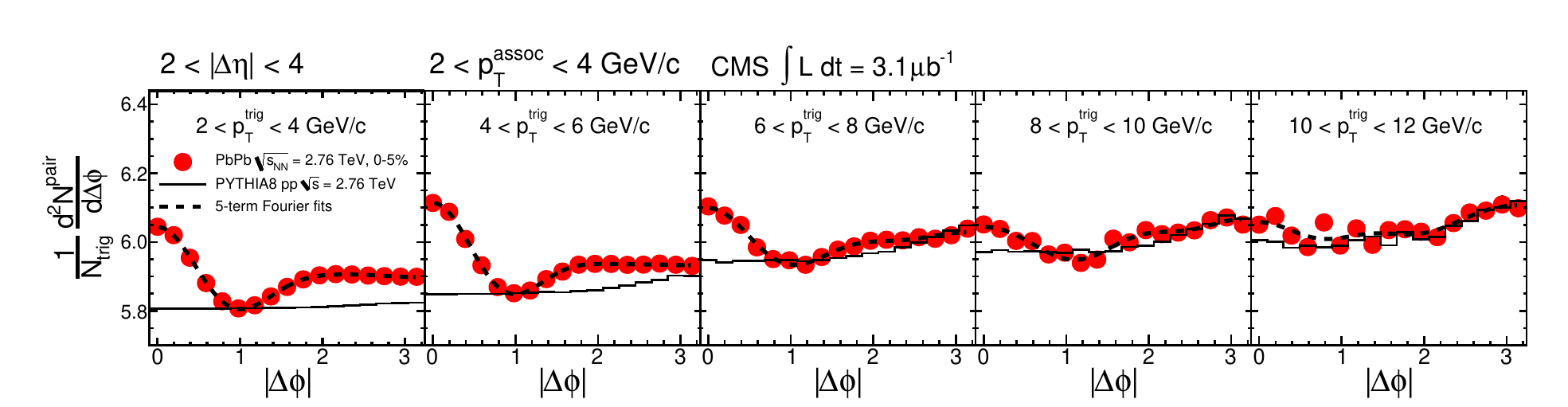}
    \caption{Long-range ($2 < |\Delta\eta| < 4$)
        per-trigger-particle associated yields of charged hadrons as a function
        of $|\Delta\phi|$
        under the same conditions as in Fig.~\ref{fig:corr1D_eta1}. The systematic
        uncertainty of 7.3\% for all data points is not shown in the plots. Dashed lines show
        the fits by the first five terms in the Fourier series, as discussed in
        Section~\ref{sec:fourier}. }
    \label{fig:corr1D_eta2}
  \end{center}
\end{figure}

\noindent
The results of extracting the 1-D $\Delta\phi$ correlations
for the 0--5\% most central PbPb collisions
are shown in Figs.~\ref{fig:corr1D_eta1}
and \ref{fig:corr1D_eta2}.
The associated yield per trigger particle in the range of
$2 < \ptass < 4$\GeVc is extracted for
five different \pttrg\ intervals (2--4, 4--6, 6--8, 8--10, and 10--$12\GeVc$) and
two ranges in $\Delta\eta$. Figure~\ref{fig:corr1D_eta1}
gives the short-range pseudorapidity result, i.e., averaged over
the region $|\Delta\eta| < 1$. Figure~\ref{fig:corr1D_eta2} shows the same comparison
for long-range correlations, i.e., averaged over the region
$2 < |\Delta\eta| < 4$.
A comparison to \PYNEW pp MC events at $\sqrt{s}=2.76$\TeV
is also shown, with a constant added to match the PbPb
results at $\Delta\phi=1$ in order to facilitate the comparison.
In this projection, only the range
$0 < \Delta\phi < \pi$ is shown, as the $\Delta\phi$ correlation function
is symmetric around $\Delta\phi = 0$ by construction.

The panels of Figs.~\ref{fig:corr1D_eta1} and \ref{fig:corr1D_eta2} show a
modification of the away-side associated yield for $|\Delta\phi| > \pi/2$
in central PbPb collisions that is not present in the \PYNEW pp MC simulation.
This modification is most pronounced for lower \pttrg\ values and
is characterized by a similarly broad distribution as seen in
lower-energy RHIC measurements, where it was attributed to jet
quenching and related phenomena~\cite{Adams:2005ph,phenix:2008cqb}.
The near-side associated yield in PbPb data includes the contributions
from both the jet-like peak and the ridge structure seen in Fig.~\ref{fig:corr2D_final}a.
Thus, it is not directly comparable to \PYNEW pp MC events, where
the ridge component is absent. A more quantitative comparison of
the jet-like component between PbPb data and \PYNEW pp MC events will
be discussed below.

For azimuthal correlations at large values of $\Delta\eta$
(Fig.~\ref{fig:corr1D_eta2}), a clear maximum
at $\Delta\phi \approx 0$ is observed, which
corresponds to the ridge structure seen in the
2-D distribution of Fig.~\ref{fig:corr2D_final}a. This
feature of the long-range azimuthal correlation function is
not present in the \PYNEW pp MC simulation for any \pttrg\ bin.
In PbPb, the height of the ridge structure decreases as
\pttrg\ increases (Fig.~\ref{fig:corr1D_eta2}) and has
largely vanished for $\pttrg \approx 10$--$12\GeVc$.
The diminishing height of the ridge with increasing \pttrg\
was not evident from previous measurements at RHIC in AuAu
collisions presumably because of lack of events at high \pt.

\subsection{Integrated Associated Yield}

The strengths of the jet peak and ridge on the near side,
as well as their dependences on $\Delta\eta$ and \pttrg,
can be quantified by the integrated associated yields.
In the presence of multiple
sources of correlations, the correlation of interest is
commonly estimated using an implementation of the zero-yield-at-minimum
(ZYAM) method~\cite{Chiu:2005ad}. However, as mentioned previously,
the possible contribution of elliptic flow is
not taken into account as it is not the dominant effect for the
most central PbPb collisions considered here.
The ZYAM method is implemented as follows.
A second-order polynomial is fitted to the $|\Delta\phi|$ distributions
in the region $0.5 < |\Delta\phi| < 1.5$. The location of the minimum of the
polynomial in this region is denoted as $\Delta\phi_{\rm ZYAM}$.
Using the position of the minimum, the associated yield is
then calculated as the integral of the $|\Delta\phi|$ distribution minus
its value at $\Delta\phi_{\rm ZYAM}$ between $|\Delta\phi| = 0$ and $\Delta\phi_{\rm ZYAM}$.
The uncertainty on the minimum level obtained by the
ZYAM procedure, combined with the uncertainty arising from the
choice of fit range in $|\Delta\phi|$, results in an uncertainty
on the absolute associated yield that is constant with a value of 0.012
over all $\Delta\eta$ and \pttrg\ bins.

\begin{figure}[thb]
  \begin{center}
    \includegraphics[width=0.6\textwidth]{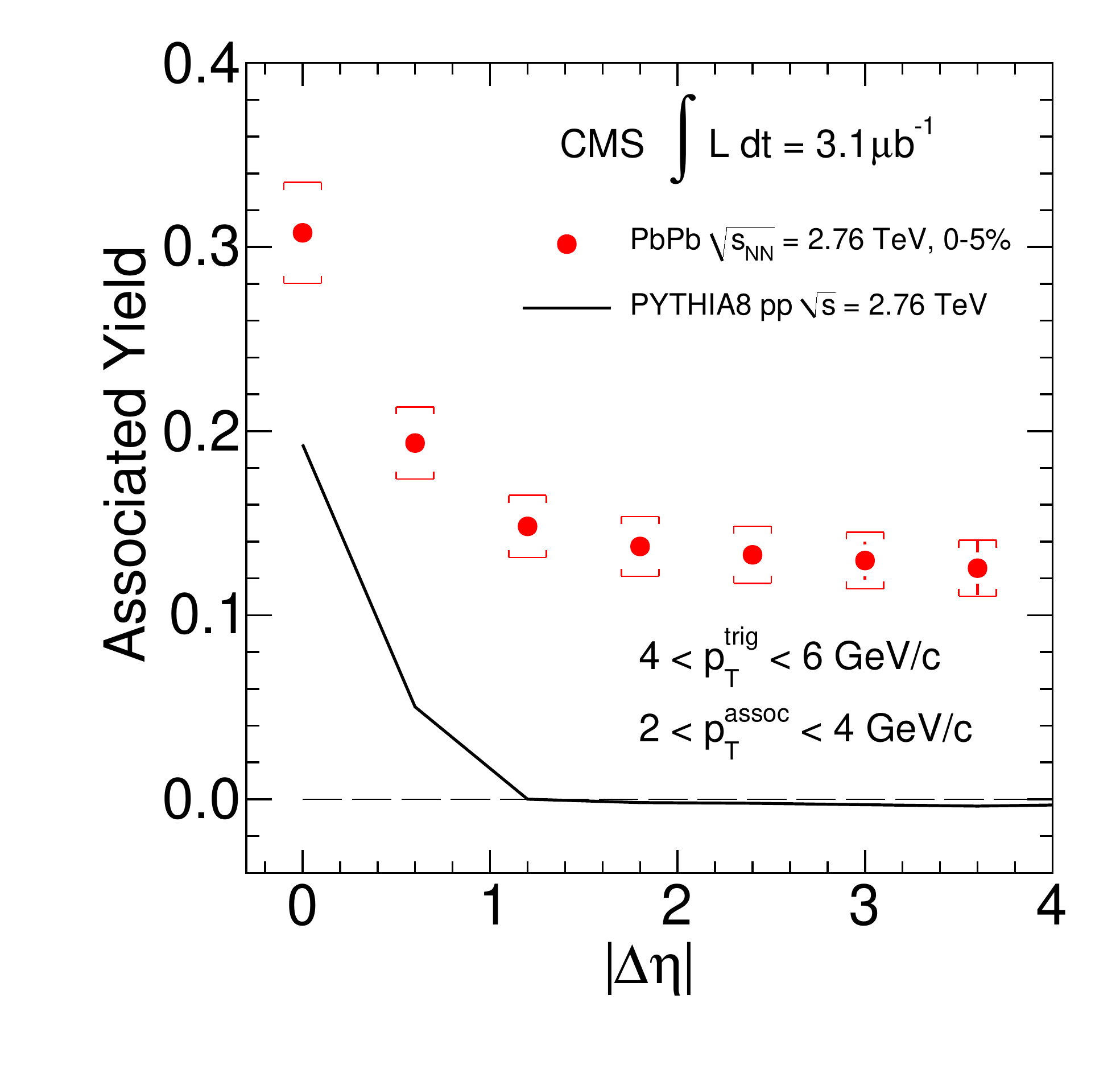}
    \caption{Integrated near-side ($|\Delta\phi| < \Delta\phi_{\rm ZYAM}$)
    associated yield for $4 < \pttrg < 6\GeVc$ and $2 < \ptass < 4\GeVc$,
    above the minimum level found by the ZYAM procedure,
    as a function of $|\Delta\eta|$ for the 0--5\% most central PbPb
    collisions at \rootsNN\ = 2.76\TeV. The error bars correspond to
    statistical uncertainties, while the brackets denote the systematic
    uncertainties. The solid line shows the prediction from the \PYNEW
    simulation of pp collisions at \roots\ = 2.76\TeV.}
    \label{fig:yield_eta}
  \end{center}
\end{figure}

\begin{figure}[thb]
  \begin{center}
    \includegraphics[width=0.85\textwidth]{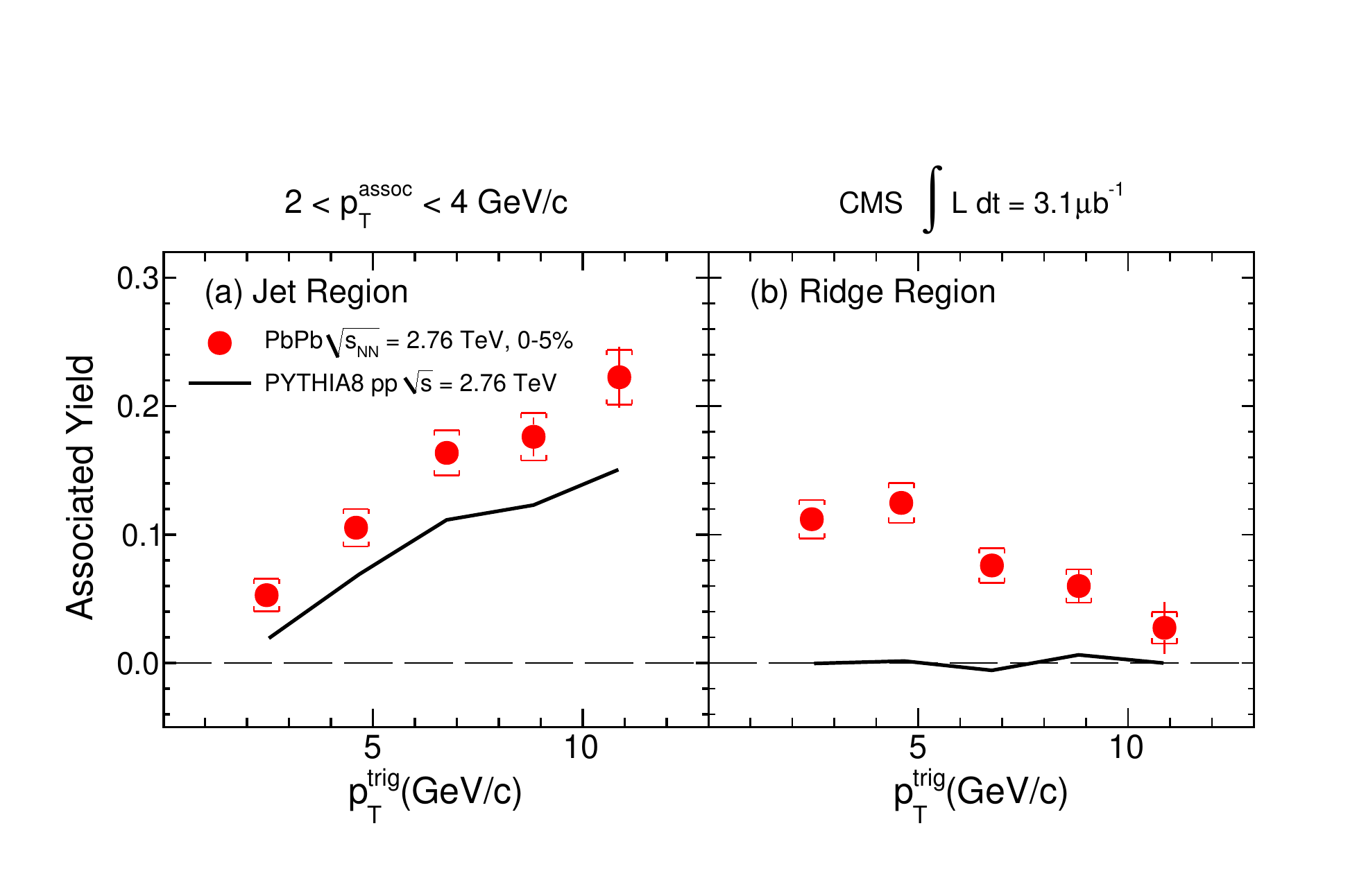}
    \caption{Integrated near-side ($|\Delta\phi| < \Delta\phi_{\rm ZYAM}$)
    associated yield above the minimum level found by the ZYAM procedure
    for (a) the short-range jet region ($|\Delta\eta| < 1$) and (b) the
    long-range ridge region ($2 < |\Delta\eta| < 4$),
    as a function of \pttrg\ in the 0--5\% most central PbPb collisions
    at \rootsNN\ = 2.76\TeV with $2 < \ptass < 4\GeVc$. The error bars correspond to
    statistical uncertainties, while the brackets around the data points denote
    the systematic uncertainties. The solid line shows the prediction from
    the \PYNEW simulation of pp collisions at \roots\ = 2.76\TeV.}
    \label{fig:yield_pt}
  \end{center}
\end{figure}

The ZYAM procedure enables the direct extraction of integrated yields of the
$\Delta\phi$-projected distributions in a well-defined manner.
Figure~\ref{fig:yield_eta} shows the resulting near-side associated yield
as a function of $|\Delta\eta|$ (in slices of 0.6 units) in both the
central PbPb collisions and \PYNEW pp MC.
The near-side associated yield for \PYNEW pp shows a strong
peak at $|\Delta\eta| = 0$, which corresponds to
the expected correlations within jets. This near-side peak
diminishes rapidly with increasing $|\Delta\eta|$.
The PbPb data also exhibit a jet-like correlation peak in the
yield for small $|\Delta\eta|$, but in contrast, the PbPb data
clearly show that the ridge extends to the highest
$|\Delta\eta|$ values measured.

Figure~\ref{fig:yield_pt} presents the integrated associated yield for
the jet region ($|\Delta\eta| < 1$) and the ridge region ($2 < |\Delta\eta| < 4$)
with $2 < \ptass < 4\GeVc$, as a function of \pttrg\ in the 0--5\% most
central PbPb collisions. The ridge-region yield is defined as the integral
of the near side in long-range $\Delta\phi$ azimuthal correlation functions
(Fig.~\ref{fig:corr1D_eta2}), while the jet-region yield is
determined by the difference between the short- and long-range near-side integral,
as the ridge is found to be approximately
constant in $\Delta\eta$ (Fig.~\ref{fig:yield_eta}).
While the jet-region yield shows an increase with \pttrg\ due to the
increasing jet transverse energy, the ridge-region yield is most prominent for
$2 < \pttrg < 6\GeVc$ and tends to drop to almost zero when \pttrg\
reaches 10--12\GeVc. The bars in Fig.~\ref{fig:yield_pt} correspond to the statistical uncertainties,
while the brackets around the data points denote the systematic uncertainties,
which are dominated by the tracking performance, as discussed earlier.
Results from the \PYNEW pp MC simulation, displayed as solid lines in Fig.~\ref{fig:yield_pt},
are consistent with zero for all \pttrg\ bins in the ridge region and have
qualitatively the same trend with \pttrg\ in the jet region as the data.
It has been seen from previous studies that the ridge is absent in both
minimum bias pp data and \PYNEW MC simulations. However, the \PYNEW program
does not fully describe the jet-like correlations observed in pp
collisions at \roots\ = 7\TeV~\cite{Khachatryan:2010gv}.
A quantitative comparison of the correlations in PbPb and pp collisions
will therefore be deferred until pp results at \roots\ = 2.76\TeV become available.

\begin{figure}[thb]
  \begin{center}
    \includegraphics[width=\linewidth]{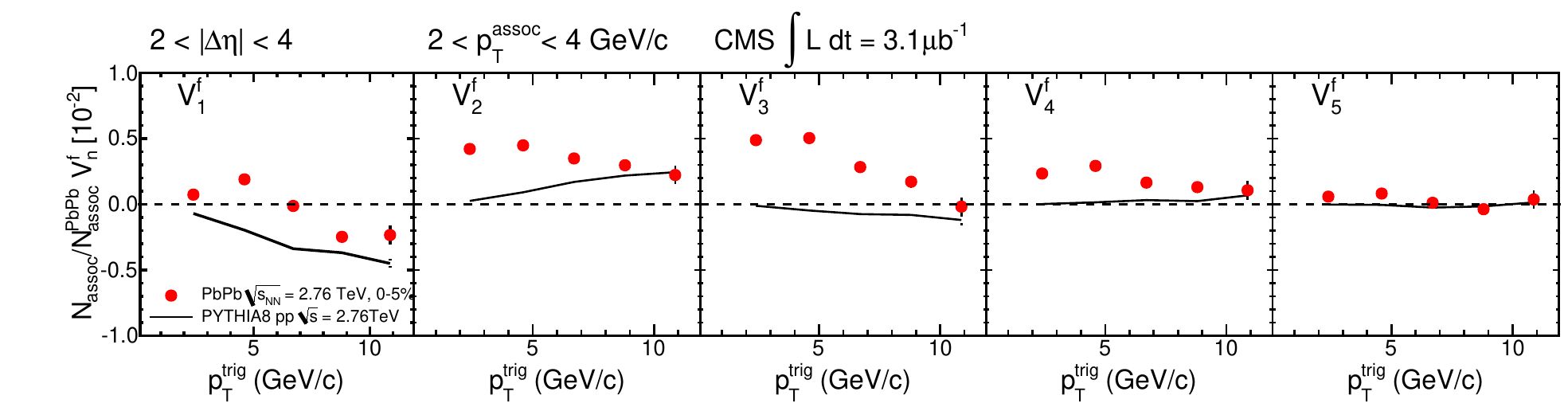}
    \caption{Fourier coefficients, $V^{\rm f}_{1}$, $V^{\rm f}_{2}$, $V^{\rm f}_{3}$,
    $V^{\rm f}_{4}$, and $V^{\rm f}_{5}$, extracted
    as functions of \pttrg\ for $2 < \ptass < 4$\GeVc for the 0--5\% most central PbPb
    collisions at \rootsNN\ = 2.76\TeV. The error bars represent statistical uncertainties only.
    The solid lines show the predictions from the \PYNEW
    simulation of pp collisions at \roots\ = 2.76\TeV.}
    \label{fig:fourier}
  \end{center}
\end{figure}

\subsection{Fourier Decomposition of $\Delta\phi$ Distributions}
\label{sec:fourier}

Motivated by recent theoretical developments in understanding the long-range
ridge effect in the context of higher-order hydrodynamic flow
induced by initial geometric fluctuations, such as the ``triangular flow''
effect~\cite{Alver:2010gr,Alver:2010dn,Schenke:2010rr,Petersen:2010cw,Xu:2010du,Teaney:2010vd},
an alternative way of quantifying the observed long-range correlations is investigated
in this paper. The 1-D $\Delta\phi$-projected distribution for $2 < |\Delta\eta| < 4$
is decomposed in a Fourier series as

\vspace{-0.2cm}
\begin{equation}
\label{fourier}
\frac{1}{N_{\rm trig}}\frac{dN^{\rm pair}}{d\Delta\phi} = \frac{N_{\rm assoc}}{2\pi} \left\{ 1+\sum\limits_{n=1}^{\infty} 2V^{\rm f}_{n} \cos (n\Delta\phi)\right\},
\end{equation}
\vspace{-0.2cm}

\noindent
where $N_{\rm assoc}$ represents the total number of hadron pairs per trigger
particle for a given $|\Delta\eta|$ range and $(\pttrg, \ptass)$ bin. The 1-D $\Delta\phi$
projections displayed in Fig.~\ref{fig:corr1D_eta2} are fitted by the first
five terms in the Fourier series, the resulting fits being shown as the dashed
lines in Fig.~\ref{fig:corr1D_eta2}. The data are well described
by the fits. Figure~\ref{fig:fourier} presents the
first five Fourier coefficients from the fit as functions of \pttrg\ for $2 < \ptass < 4$\GeVc
for the 0--5\% most central PbPb collisions and for the \PYNEW pp MC simulation. The
\PYNEW results are scaled by the ratio of $N_{\rm assoc}^{\rm \PYNEW}/N_{\rm assoc}^{\rm PbPb}$
in order to remove the trivial effect of the multiplicity dependence. The error bars are
statistical only, while the systematic uncertainties are found to be negligible because
the Fourier coefficients characterize the overall shape of the correlation
functions, and thus are not sensitive to the absolute scale.
The coefficients from the fit are also found to be largely independent of each other
(correlation coefficients typical below 5\%).

All the Fourier coefficients found from fitting the PbPb data show
a similar dependence on \pttrg\ except for the $V^{\rm f}_{1}$ term, which
contains an additional negative contribution that grows
toward higher \pttrg. This negative $V^{\rm f}_{1}$ component is consistent with a contribution
from momentum conservation or back-to-back dijets~\cite{Luzum:2010fb}.
If the observed correlation is purely driven by the single-particle azimuthal
anisotropy arising from the hydrodynamic expansion of the medium~\cite{Voloshin:1994mz},
the extracted $V^{\rm f}_{n}$ components would be
related to the flow coefficients $v_{n}$ (i.e., $v_{2}$ for
anisotropic elliptic flow) via
$V^{\rm f}_{n} \sim v^{\rm trig}_{n} \times v^{\rm assoc}_{n}$,
where $v^{\rm trig}_{n}$ and $v^{\rm assoc}_{n}$ are the flow coefficients for
the trigger and associated particles~\cite{Alver:2010gr}.
The flow coefficients, and especially the higher-order terms, are sensitive
to the initial conditions and viscosity of the hot, dense medium~\cite{Alver:2010dn,Staig:2010pn}.
This Fourier analysis serves as an alternative way of quantifying the azimuthal
correlation functions in heavy-ion collisions and potentially provides new insights in
understanding the origin of the ridge effect in these collisions.
Analysis of larger data samples in terms of \pttrg\ and \ptass, as well as
centrality, would allow a more detailed comparison to
theoretical calculations of hydrodynamics, for which
the Fourier components provide a concise description of the data.

\section{Summary}
\label{sec:conclusions}
The CMS detector at the LHC has been used to measure
angular correlations between charged particles
in $\Delta \eta$ and $\Delta \phi$ up to $|\Delta \eta| \approx 4$
and over the full range of $\Delta \phi$ in the 0--5\% most central
PbPb collisions at \rootsNN\ = 2.76\TeV. This is the first study of
long-range azimuthal correlations over a large difference in
pseudorapidity in PbPb interactions at the LHC energy. The extracted
2-D associated yield distributions show a variety of characteristic features
in heavy-ion collisions that are not present in minimum bias pp interactions.
Short- and long-range azimuthal correlations have been studied as a function
of the transverse momentum of the trigger particles. The
observed long-range ridge-like structure for approximately equal azimuthal angles
($\Delta\phi \approx 0$) is most evident in the intermediate transverse momentum
range, $2 < \pttrg < 6$\GeVc, and decreases to almost zero for \pttrg\ above 10--12\GeVc.
A qualitatively similar dependence of the ridge on transverse momentum has
also been observed in high-multiplicity pp events, indicating a potentially similar
physical origin of the effect. A Fourier decomposition of the 1-D $\Delta\phi$-projected
correlation functions in the ridge region ($2 < |\Delta\eta| < 4$) has been presented.
This alternative way of quantifying the correlation data provides valuable
information to test a wide range of theoretical models, including recent
hydrodynamic calculations of higher-order Fourier components.
The very broad solid-angle coverage of the CMS detector and the
statistical accuracy of the sample analyzed in this paper provide
significantly improved observations of short- and long-range particle
correlations over previously available measurements.

\section*{Acknowledgments}
We wish to congratulate our colleagues in the CERN accelerator departments for the excellent performance of the LHC machine. We thank the technical and administrative staff at CERN and other CMS institutes, and acknowledge support from: FMSR (Austria); FNRS and FWO (Belgium); CNPq, CAPES, FAPERJ, and FAPESP (Brazil); MES (Bulgaria); CERN; CAS, MoST, and NSFC (China); COLCIENCIAS (Colombia); MSES (Croatia); RPF (Cyprus); Academy of Sciences and NICPB (Estonia); Academy of Finland, MEC, and HIP (Finland); CEA and CNRS/IN2P3 (France); BMBF, DFG, and HGF (Germany); GSRT (Greece); OTKA and NKTH (Hungary); DAE and DST (India); IPM (Iran); SFI (Ireland); INFN (Italy); NRF and WCU (Korea); LAS (Lithuania); CINVESTAV, CONACYT, SEP, and UASLP-FAI (Mexico); MSI (New Zealand); PAEC (Pakistan); SCSR (Poland); FCT (Portugal); JINR (Armenia, Belarus, Georgia, Ukraine, Uzbekistan); MST and MAE (Russia); MSTD (Serbia); MICINN and CPAN (Spain); Swiss Funding Agencies (Switzerland); NSC (Taipei); TUBITAK and TAEK (Turkey); STFC (United Kingdom); DOE and NSF (USA).

Individuals have received support from the Marie-Curie programme and the European Research Council (European Union); the Leventis Foundation; the A. P. Sloan Foundation; the Alexander von Humboldt Foundation; the Associazione per lo Sviluppo Scientifico e Tecnologico del Piemonte (Italy); the Belgian Federal Science Policy Office; the Fonds pour la Formation \`a la Recherche dans l'Industrie et dans l'Agriculture (FRIA-Belgium); the Agentschap voor Innovatie door Wetenschap en Technologie (IWT-Belgium); and the Council of Science and Industrial Research, India.

\bibliography{auto_generated}   
\clearpage \cleardoublepage\appendix\section{The CMS Collaboration \label{app:collab}}\begin{sloppypar}\hyphenpenalty=5000\widowpenalty=500\clubpenalty=5000\textbf{Yerevan Physics Institute,  Yerevan,  Armenia}\\*[0pt]
S.~Chatrchyan, V.~Khachatryan, A.M.~Sirunyan, A.~Tumasyan
\vskip\cmsinstskip
\textbf{Institut f\"{u}r Hochenergiephysik der OeAW,  Wien,  Austria}\\*[0pt]
W.~Adam, T.~Bergauer, M.~Dragicevic, J.~Er\"{o}, C.~Fabjan, M.~Friedl, R.~Fr\"{u}hwirth, V.M.~Ghete, J.~Hammer\cmsAuthorMark{1}, S.~H\"{a}nsel, M.~Hoch, N.~H\"{o}rmann, J.~Hrubec, M.~Jeitler, W.~Kiesenhofer, M.~Krammer, D.~Liko, I.~Mikulec, M.~Pernicka, H.~Rohringer, R.~Sch\"{o}fbeck, J.~Strauss, A.~Taurok, F.~Teischinger, P.~Wagner, W.~Waltenberger, G.~Walzel, E.~Widl, C.-E.~Wulz
% ...
\vskip\cmsinstskip
% ...
1:~~Also at CERN, European Organization for Nuclear Research, Geneva, Switzerland\\
% ...

\end{sloppypar}
\end{document}